\documentclass[pre,aps,twocolumn,superscriptaddress]{revtex4-1}

\usepackage{graphicx}
\usepackage{amssymb,amsfonts,amsmath}
\usepackage{color}
\usepackage{ulem}
\usepackage[hidelinks]{hyperref}
\usepackage{bbm}
\usepackage{bbold}

\usepackage{tikz}
\usetikzlibrary{shapes}
\usetikzlibrary{patterns}
\usetikzlibrary{angles, quotes}

\newcommand{\rv}{{\mathbf r}}

\newcommand{\Tr}{{\rm Tr}\,}
\newcommand{\e}{{\rm e}}

\newcommand{\pv}{{\bf p}}

\newcommand{\msphantom}[1]{$\ldots$}

\newcommand{\eqr}[1]{Eq.~\eqref{#1}}

\newcommand{\mydelete}[1]{{}}

\newcommand{\rmint}{{\rm int}}

\newcommand{\rmexc}{{\rm exc}}
\newcommand{\rmext}{{\rm ext}}

\newcommand{\rmid}{{\rm id}}

\newcommand{\rmr}{{\rm ref}}
\newcommand{\calV}{{\cal V}}

\newcommand{\cov}{{\rm cov}}

\begin{document}

\title{Metadensity functional theory for classical fluids:
Extracting the pair potential}

\author{Stefanie M. Kampa}
\affiliation{Theoretische Physik II, Physikalisches Institut, 
  Universit{\"a}t Bayreuth, D-95447 Bayreuth, Germany}

\author{Florian Samm\"uller}
\affiliation{Theoretische Physik II, Physikalisches Institut, 
  Universit{\"a}t Bayreuth, D-95447 Bayreuth, Germany}

\author{Matthias Schmidt}
\affiliation{Theoretische Physik II, Physikalisches Institut, 
  Universit{\"a}t Bayreuth, D-95447 Bayreuth, Germany}
\email{Matthias.Schmidt@uni-bayreuth.de}

\author{Robert Evans}
\affiliation{H. H. Wills Physics Laboratory, 
   University of Bristol, Royal Fort, Bristol BS8 1TL, United Kingdom}

\date{11 November 2024, revised version: 8 January 2025}

\begin{abstract}
 The excess free energy functional of classical density functional
 theory depends upon the type of fluid model, specifically on the
 choice of (pair) potential, is unknown in general, and is
 approximated reliably only in special cases. We present a machine
 learning scheme for training a neural network that acts as a generic
 metadensity functional for truncated but otherwise arbitrary pair
 potentials.  Automatic differentiation and neural functional calculus
 then yield, for one-dimensional fluids, accurate predictions for
 inhomogeneous states and immediate access to the pair distribution
 function.  The approach provides a means of addressing a fundamental
 problem in the physics of liquids, and for soft matter design: `How
 best to invert structural data to obtain the pair potential?'
\end{abstract}

\maketitle

Classical density functional theory (DFT) is a powerful framework for
investigating the equilibrium structure and thermodynamics of bulk and
spatially inhomogeneous liquids and more general soft matter systems
\cite{evans1979, evans1992, hansen2013, evans2016,
  schmidt2022rmp}. Rosenfeld's fundamental measure theory for hard
spheres proved pivotal to the field, due to its high accuracy and its
beautiful and intriguing geometrical structure~\cite{rosenfeld1988,
  rosenfeld1989, tarazona2000, roth2010}. The treatment of
longer-ranged interparticle attraction, acting on top of short-ranged
repulsion, is typically based on a simple additive mean-field
contribution to the (accurate) hard sphere free energy density
functional \cite{evans1992, hansen2013}.  Despite its status as a
workhorse of DFT, the predictions of the `standard mean field
approach' rarely allow for direct quantitative comparison to
simulation results.

To overcome limitations of classical DFT approximations a range of
recent studies addressed the possibility to apply machine learning
techniques \cite{teixera2014, lin2019ml, lin2020ml, cats2022ml,
  qiao2020, yatsyshin2022, malpica-morales2023, fang2022,
  dijkman2024ml, kelley2024ml, delasheras2023perspective}.  The neural
functional theory \cite{sammueller2023neural, sammueller2023whyNeural,
  sammueller2023neuralTutorial, sammueller2024hyperDFT,
  sammueller2024whyhyperDFT, sammueller2024pairmatching,
  sammueller2024attraction, bui2024neuralrpm}, based on representing
the one-body direct correlation functional, the first derivative of
the excess free energy functional, by a neural network, has proved
very successful. The approach allows for extensive use of functional
integration and differentiation techniques \cite{sammueller2023neural,
  sammueller2023whyNeural, sammueller2024hyperDFT,
  sammueller2024whyhyperDFT, sammueller2024pairmatching,
  sammueller2023neuralTutorial, sammueller2024attraction,
  bui2024neuralrpm} using automatic
differentiation~\cite{baydin2018autodiff, stierle2024autodiff}.
Neural functionals were trained for specific interparticle interaction
potentials, including three-dimensional hard spheres
\cite{sammueller2023neural}, one-dimensional hard
\cite{sammueller2023whyNeural} and attractive rods
\cite{sammueller2024hyperDFT, sammueller2024whyhyperDFT},
three-dimensional supercritical~\cite{sammueller2023neural,
  sammueller2024pairmatching, sammueller2024attraction} and
subcritical~\cite{sammueller2024attraction} Lennard-Jones (LJ) fluids,
and ionic fluids~\cite{bui2024neuralrpm}.

At the core of these successful machine learning applications
\cite{sammueller2023neural, sammueller2023whyNeural,
  sammueller2024hyperDFT, sammueller2024whyhyperDFT,
  sammueller2024pairmatching, sammueller2023neuralTutorial,
  sammueller2024attraction, bui2024neuralrpm} lie the formally exact
functional relationships of DFT \cite{evans1979}. In particular the
density profile $\rho(\rv)$ determines the one-body direct correlation
function $c_1(\rv)$. Of course, the interparticle interactions and the
statepoint need to be known. For model fluids where particles interact
solely with a pair potential $\phi(r)$, with~$r$ denoting the
interparticle distance, Henderson's
theorem~\cite{henderson1974uniqueness} states that knowledge of the
pair distribution function~$g(r)$ at a single statepoint is sufficient
to determine the pair potential $\phi(r)$ up to a constant.  Whilst
being a foundational problem
\cite{johnson1964,levesque1985,enderby1988, evans1990reverseMC,
  percus1986}, much current research is devoted to exploring the
practicalities and consequences, e.g., for reverse Monte
Carlo~\cite{mcgreevy2001, evans1990reverseMC}, variational methods
\cite{frommer2019, frommer2024}, machine learning
\cite{moradzadeh2021deeplearning, sammueller2024pairmatching,
  toth2005}, sensitivity \cite{wang2020jcp, wang2022pre}, design tasks
\cite{jadrich2015, wang2024jcp}, quasiuniversality~\cite{dyre2016},
and microscopy in colloidal systems \cite{stones2019prl}.

Here we demonstrate the feasibility of neural functional training
based on local learning of the simultaneous functional dependencies on
both the density profile~$\rho(\rv)$ and on the thermally scaled pair
potential~$\beta \phi(r)$, where $\beta=1/(k_BT)$, with the Boltzmann
constant $k_B$ and temperature $T$. The increase in demand of training
data generated from simulations of spatially inhomogeneous systems
with varying pair potentials is only very moderate, given the
significant increase both in functional complexity and applicability
to general form of~$\phi(r)$ and the free choice of the
temperature~$T$. As the functional dependence on $\phi(r)$ is often
merely treated implicitly, we refer to explicit and accurate
functionals of {\it both} density and pair potential as {\it
  metadensity} functionals. We apply and test the framework for
general one-dimensional interacting fluid models after first
describing the underlying theoretical structure.

We consider systems of $N$ particles with identical mass $m$, position
coordinates $\rv_i$ and momenta $\pv_i$, where the index $i$ labels
the particles with $i=1,\ldots,N$. The Hamiltonian~$H$ is taken to
consist of a sum of an intrinsic and an external contribution, $H =
H_\rmint + \sum_{i=1}^N V_\rmext(\rv_i)$, where $V_\rmext(\rv)$ is an
external potential that acts at (generic) position $\rv$. The
intrinsic part $H_\rmint$ of the Hamiltonian is the sum of kinetic and
interparticle interaction energy, which we restrict to pair
contributions only, such that
\begin{align}
  H_\rmint &= \sum_{i=1}^{N} \frac{\pv_i^2}{2m} 
  + \frac{1}{2}
  \sum_{i=1}^N\sum_{j=1, j\neq i}^N \phi(|\rv_i-\rv_j|),
  \label{EQHint}
\end{align}
and we comment on more general forms of interparticle potentials
$u(\rv_1,\ldots,\rv_N)$ below.

The definition of the intrinsic free energy functional can be based on
Levy's constrained search \cite{levy1979, dwandaru2011} or on the
standard Mermin argument \cite{evans1979, mermin1965, hansen2013}.
The intrinsic free energy functional consists of an ideal and an
excess part according to:
\begin{align}
  F[\rho,\beta\phi] &= 
  F_\rmid[\rho] + F_\rmexc[\rho,{\beta}\phi],
\end{align}
where $\beta F_\rmid[\rho] = \int d\rv
\rho(\rv)[\ln(\rho(\rv)\Lambda^d)-1]$ with dimensionality $d$ and
thermal de Broglie wavelength $\Lambda$; for convenience we set
$\Lambda=1$ in the following. The excess free energy functional
$F_\rmexc[\rho,\beta\phi]$ accounts for the effects of the
interparticle interactions via its functional dependence on
$\beta\phi(r)$. Thermal scaling incorporates fully the dependence on
temperature $T$, such that there remains no hidden temperature
dependence in $\beta F[\rho,\beta\phi]$, as follows, e.g., from the
definition via Levy search, $\beta F[\rho,\beta\phi]=\min_{f\to\rho}
\Tr f(\beta H_\rmint + \ln f)$, where $\Tr$ denotes the grand
canonical trace, and $f$ the many-body trial distribution function
\cite{levy1979, dwandaru2011}. The one-body direct correlation
functional follows via functional differentiation \cite{evans1979,
  evans1992, hansen2013, evans2016, schmidt2022rmp}:
\begin{align}
  c_1(\rv;[\rho,\beta \phi]) &= 
  -\frac{\delta \beta F_\rmexc[\rho,\beta \phi]}{\delta\rho(\rv)}.
  \label{EQc1AsDerivative}
\end{align}

\begin{figure}[!t]
  \vspace{1mm}
  \includegraphics[width=.9\columnwidth]{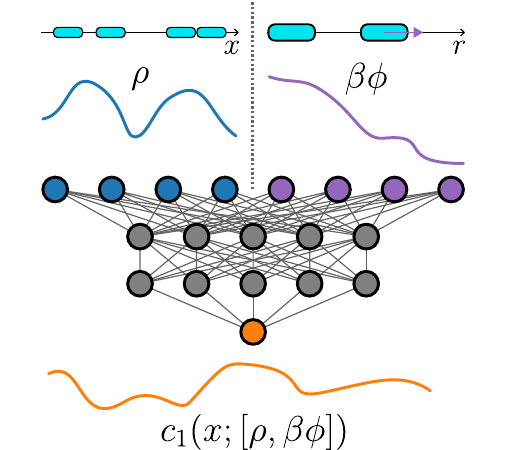}
  \caption{Training strategy of the neural metadensity functional.
    Grand canonical Monte Carlo simulations yield training data for
    supervised machine learning. The simulations are carried out with
    randomized scaled external potentials $\beta V_\rmext(x)$ and
    randomized scaled chemical potentials $\beta\mu$. Results for the
    density profiles $\rho(x)$ are obtained for randomized scaled
    interparticle potentials $\beta\phi(r)$. These data sets are used
    to train a neural one-body direct correlation metadensity
    functional $c_1(x;[\rho,\beta\phi])$ that renders the simultaneous
    functional dependence on both fields $\rho(x)$ and $\beta\phi(r)$
    operational. Here $x$ is the one-dimensional position coordinate
    and $r$ is interparticle distance. We have used 12089 simulation
    runs and truncated $\rho(x)$ symmetrically around each grid point
    at $\pm 2\sigma$. We restrict ourselves to $\beta\phi(r)$ that
    vanish beyond the length $r_c=1.5\sigma$ of its input grid, where
    $\sigma$ is the lengthscale and $0.01\sigma$ is the grid spacing
    \cite{kampa2024metaData}.}
\label{FIG1}
\end{figure}

We aim to transcend existing neural functional methods
\cite{sammueller2023neural, sammueller2023whyNeural,
  sammueller2023neuralTutorial, sammueller2024hyperDFT,
  sammueller2024whyhyperDFT, sammueller2024pairmatching,
  sammueller2023neuralTutorial, sammueller2024attraction,
  bui2024neuralrpm} and base our supervised local machine learning on
simulation data not only generated by varying the form of the external
potential $V_\rmext(\rv)$, but also by varying the pair potential
$\phi(r)$.  For any given training system (given $\phi(r)$) sampling
in grand canonical Monte Carlo simulations yields the equilibrium
density profile $\rho(\rv)$ at prescribed~$T$ and chemical potential
$\mu$. The resulting form of the one-body direct correlation function,
intended to serve as a reference for the neural training, then follows
as
\begin{align}
  c_1^\rmr(\rv) &= 
  \ln\rho(\rv) + \beta V_\rmext(\rv) - \beta\mu.
  \label{EQc1reference}
\end{align}
Note that each term on the right hand side is known, as was exploited
in previous work \cite{sammueller2023neural, sammueller2023whyNeural,
  sammueller2023neuralTutorial, sammueller2024hyperDFT,
  sammueller2024whyhyperDFT, sammueller2024pairmatching,
  sammueller2024attraction, bui2024neuralrpm}, and that
\eqr{EQc1reference} is valid for all interparticle potentials.

\begin{figure*}[!t]
  \vspace{1mm}
  \includegraphics[width=.99\textwidth]{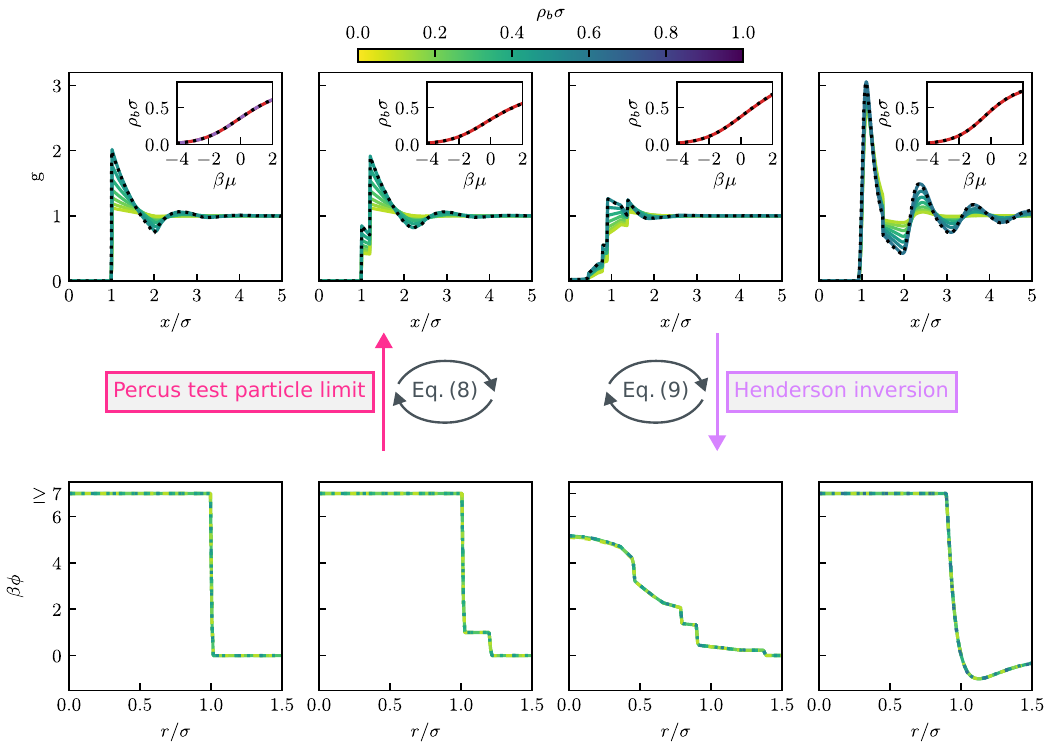}
  \caption{Pair structure and Henderson design via the neural
    metadensity functional. Results are shown for the pair
    distribution function $g(r)$ (first row) and corresponding scaled
    pair potential $\beta\phi(r)$ (second row) for different values of
    scaled bulk density $\rho_b\sigma$ (colour bar) and for four
    different pair potentials $\beta\phi(r)$ in one dimension: i)~hard
    rods (first column), ii)~square shoulder repulsion (second
    column), iii)~random penetrable repulsion (third column), and
    iv)~LJ particles (fourth column). Results for $g(r)$ from test
    particle minimization~\eqref{EQEulerLagrangeTestParticle} with
    $\beta\phi(r)$ fixed and $\beta\mu=-2,-1.5,1,-0.5,0,0.5,1$ (yellow
    to green lines) are numerically identical to the simulation
    reference (black dots, shown for $\beta\mu=1$).  The variable $r$
    denotes one-dimensional relative distance. Insets: The neural
    equations of state $\rho_b\sigma$ versus $\beta\mu$ (red lines)
    agree with reference simulation data (black dots) and for hard
    rods with the exact solution (dashed violet line, first panel); we
    set the de Broglie wavelength $\Lambda=\sigma$ in the ideal gas
    contribution.  Using solely $g(r), \rho_b, \beta$ as input,
    Henderson density functional inversion~\eqref{EQhendersonOracle}
    allows one to reconstruct $\beta\phi(r)$ consistently (overlapping
    dashed coloured lines) and with high accuracy across the entire
    range of densities considered (colour bar); we take values
    $\beta\phi(r) \geq 7$ to represent numerically hard cores.}
\label{FIG2}
\end{figure*}

This data acquisition process is repeated several thousand times for
different randomized forms of the scaled external potential $\beta
V_\rmext(\rv)$, randomized values of the scaled chemical potential
$\beta \mu$, and, crucially, also randomized forms of the scaled pair
potential $\beta \phi(r)$. Apart from the last ingredient, the
training follows that in Refs.~\cite{sammueller2023neural,
  sammueller2023whyNeural, sammueller2023neuralTutorial,
  sammueller2024pairmatching, sammueller2024attraction,
  bui2024neuralrpm}. Now we have training data for the construction of
the neural metadirect correlation functional $c_1(\rv;[\rho,\beta
  \phi])$, which depends on both the density profile $\rho(\rv)$ and
on the form of the scaled pair interaction potential~$\beta
\phi(r)$. Specifically, the training aims to achieve equality across
the entire training data set:
\begin{align}
  c_1(\rv;[\rho,\beta\phi]) &= c_1^\rmr(\rv),
  \label{EQc1targetEquality}
\end{align}
where the left hand side indicates the neural network output and the
right hand side the training reference \eqref{EQc1reference}. We
represent $c_1(\rv;[\rho,\beta\phi])$ by a simple multi-layer
perceptron with five hidden layers; see Fig.~\ref{FIG1} for an
illustration in one-dimensional geometry and
Refs.~\cite{sammueller2024pairmatching, dijkman2024ml} for possible
further architectures.

Ready access to $c_1(\rv,[\rho,\beta\phi])$ allows one to obtain the
excess free energy functional via functional line integration $-\beta
F_\rmexc[\rho,\beta\phi] = \int {\cal D}[\rho]
c_1(\rv;[\rho,\beta\phi])$. In practice a parameterization with simple
scaling $a \rho(\rv)$ by a parameter $0\leq a \leq 1$ allows one to
numerically integrate according to $-\beta F_\rmexc[\rho,\beta\phi] =
\int d\rv \rho(\rv) \int_0^1 da c_1(\rv;[a \rho,\beta\phi])$.  The
required equilibrium density profile $\rho(\rv) $ is obtained from
self-consistent solution of the standard Euler-Lagrange equation of
DFT \cite{evans1979, evans1992, hansen2013, evans2016,
  schmidt2022rmp}, as follows formally from combining
Eqs.~\eqref{EQc1reference} and \eqref{EQc1targetEquality}:
$c_1(\rv;[\rho,\beta\phi]) = \ln\rho(\rv) + \beta V_\rmext(\rv) -
\beta \mu$, where we have kept the functional dependence on the scaled
pair potential $\beta\phi(r)$ explicit in the notation.
Exponentiating and re-arranging yields
\begin{align}
  \rho(\rv) =
  \exp(c_1(\rv;[\rho,\beta\phi]) - \beta V_\rmext(\rv)+\beta \mu),
  \label{EQEulerLagrangeMeta}
\end{align}
which suits the application of iterative solution methods.
\begin{figure*}[!t]
  \vspace{1mm}
  \includegraphics[width=.99\textwidth]{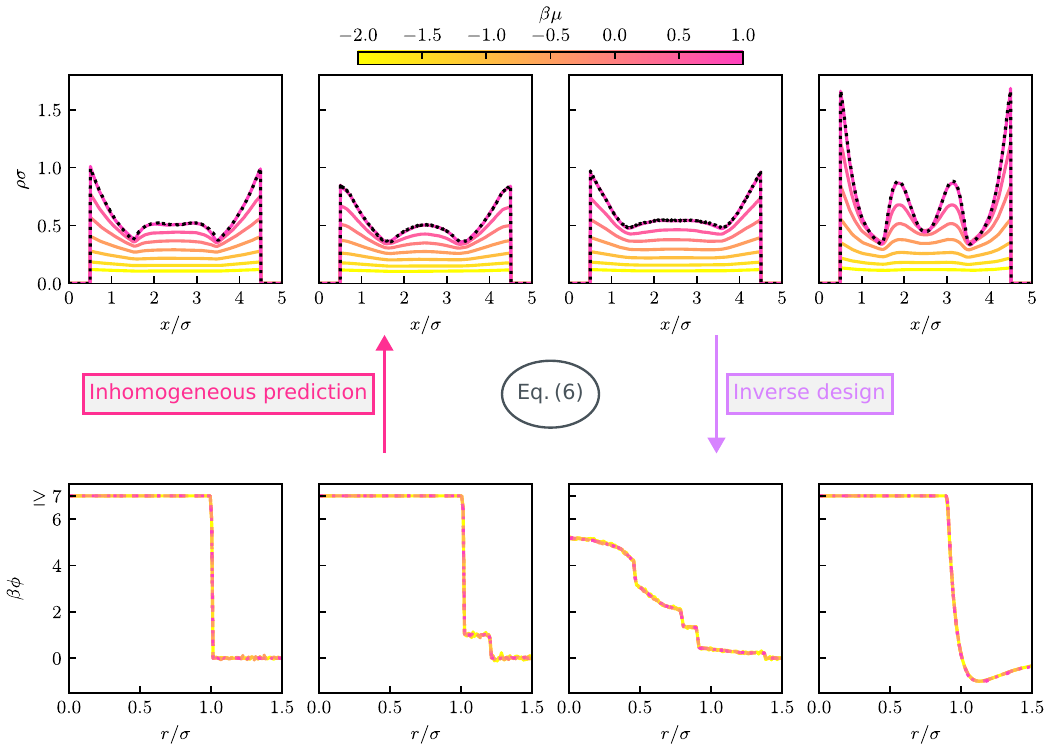}
  \caption{Metadensity functional application for confinement between
    two hard walls located at $x/\sigma= 0.5$ and 4.5 at varying
    values of the scaled chemical potential $\beta\mu$ (colour
    bar). Results are shown for scaled one-body density profiles
    $\rho(x)\sigma$ (first row, lines) from density functional
    minimization \eqref{EQEulerLagrangeMeta} for each of the four
    representative pair potentials $\beta\phi(r)$ of Fig.~\ref{FIG2}
    fixed (second row). Simulation results (black dotted lines) are
    shown for $\beta\mu=1$. Setting the density profile $\rho(x)$ as
    the target and keeping the hard walls unchanged, inhomogeneous
    inverse pair potential design aims to find the corresponding,
    as-yet unknown, pair potential $\beta\phi(r)$, where $r$ indicates
    one-dimensional distance. For each system the results for
    $\beta\phi(r)$ (second row), obtained from
    \eqr{EQEulerLagrangeMeta} for each of the seven values of $\beta
    \mu$ separately, coincide numerically with each other as well as
    with the reference, upto very small artifacts--zooming shows weak
    oscillations in overlapping dashed coloured lines. This
    demonstrates successful pair potential reconstruction and inverse
    design.}
\label{FIG3}
\end{figure*}

We specialize first to bulk fluids where $V_\rmext(\rv)=0$ and where
we expect a spatially homogeneous bulk fluid with constant density,
$\rho(\rv)=\rho_b=\rm const$. The one-body direct correlation
functional is then independent of position $\rv$ and directly related
to the excess chemical potential $\mu_\rmexc$:
\begin{align}
  c_1(\rv;[\rho_b,\beta\phi]) &= -\beta\mu_\rmexc = \ln\rho_b - \beta\mu,
  \label{EQchemicalPotentialBalance}
\end{align}
which follows directly from \eqr{EQEulerLagrangeMeta}.

Percus' test particle limit \cite{percus1962} allows one to access the
bulk pair structure. The external potential is chosen to be identical
to the pair potential of the fluid, $V_\rmext(\rv)=\phi(r)$. The
corresponding test particle density profile is $\rho(\rv)=\rho_b
g(r)$, where $g(r)$ is the standard pair distribution function
\cite{hansen2013, percus1962}.  The general Euler-Lagrange equation
\eqref{EQEulerLagrangeMeta} then attains test particle form,
\begin{align}
  \rho_b g(r) &= \exp(c_1(r;[\rho_b g,\beta\phi]) 
  - \beta \phi(r) + \beta\mu),
  \label{EQEulerLagrangeTestParticle}
\end{align}
which applies to any form of $\phi(r)$.

We describe results for four very different {\it one-dimensional}
fluids, see Fig.~\ref{FIG2}, second row. Note that~$r$ is then
replaced by $x$ in \eqr{EQEulerLagrangeTestParticle}. The models are
representative: hard rods \cite{percus1976, robledo1981,
  sammueller2023whyNeural}, square shoulder particles
\cite{montero2019}, random penetrable repulsive particles, and LJ
particles.

{\it i) Predicting pair structure: $g(r)$.}--We first take the
Hamiltonian \eqref{EQHint} to be known. The given~$\phi(r)$ can then
be used explicitly in the test particle
equation~\eqref{EQEulerLagrangeTestParticle}.  Self-consistent
solution is efficiently obtained for given values of $\beta, \mu$ via
Picard iteration, which delivers results directly for the pair
distribution function~$g(r)$.  This density functional setup of
Percus' test particle limit applies to any $\phi(r)$ using the same
universal metadensity functional $c_1(\rv;[\rho,\beta\phi])$, as is
demonstrated in Fig.~\ref{FIG2} for the four one-dimensional model
fluids. Note the pronounced effect of interparticle attraction (fourth
column) on the degree of structuring in $g(r)$.  The neural equation
of state $\rho_b(\mu)$ follows from \eqr{EQchemicalPotentialBalance},
see insets in Fig.~\ref{FIG2}. The results are numerically identical
to reference data obtained from grand canonical Monte Carlo
simulations and, for hard rods, to the exact solution \cite{tonks1936,
  percus1976}, $\beta\mu_\rmexc= -\ln(1-\eta) + \eta/(1-\eta)$ with
$\eta=\rho_b\sigma$.

{\it ii) Henderson inversion yields pair potential $\phi(r)$}.--We
secondly consider the situation where $g(r)$ is known.  We take
$\rho_b, \beta$ as the control parameters and obtain $\mu$ from
\eqr{EQchemicalPotentialBalance}. The task is to find $\phi(r)$ that
generates this given $g(r)$, which constitutes Henderson's inversion
problem~\cite{henderson1974uniqueness}; see
Refs.~\cite{dijkman2024ml,johnson1964,levesque1985,enderby1988,
  evans1990reverseMC, percus1986, mcgreevy2001, evans1990reverseMC,
  frommer2019, frommer2024, moradzadeh2021deeplearning,
  sammueller2024pairmatching, toth2005, wang2020jcp, wang2022pre,
  jadrich2015, wang2024jcp, dyre2016, stones2019prl}. We render the
formal inversion operational by taking the logarithm of
\eqr{EQEulerLagrangeTestParticle} and solving for the explicit
occurrence of the pair potential:
\begin{align}
  \beta\phi(r)   &=
  c_1(r;[\rho_b g,\beta\phi])  + \beta\mu -  \ln[\rho_bg(r)].
  \label{EQhendersonOracle}
\end{align}  
Equation \eqref{EQhendersonOracle} constitutes a self-consistency
relation for determining $\phi(r)$. Using Picard iteration, we obtain
remarkably consistent results, as demonstrated in Fig.~\ref{FIG2}. On
the scale of the figure it is hardly possible to find differences from
the (original) pair potential.

{\it iii) Inhomogeneous metadensity functional application.}--Our
method is designed for general density functional setups, where an
external potential $V_\rmext(\rv)$ is specified and one aims to
calculate the emerging inhomogeneous density profile $\rho(\rv)$ for a
given model fluid. This implies standard solution of the
Euler-Lagrange equation~\eqref{EQEulerLagrangeMeta}, keeping $\phi(r)$
fixed. Figure~\ref{FIG3} illustrates application of this strategy to
confinement between two hard walls; this situation was not part of the
training set. Note the highly structured density profiles for the LJ
fluid in the fourth column. We emphasize that the four fluid models
are mere representative examples and that all results stem from the
same neural metadensity functional $c_1(\rv,[\rho,\beta\phi])$. In
situations where the full functional dependence is not required, one
can work with parametric dependence on pair potential parameters, see
Appendix~A for application to square-shoulder particles and for
discussion of many-body interparticle potentials.

{\it iv) Soft matter inverse design.}--The flexible dependence on the
pair potential inherent in \eqr{EQEulerLagrangeMeta} allows one to go
beyond standard density functional tasks and address design problems,
similar to Henderson inversion. We hence set the target as the
inhomogeneous one-body density profile. This introduces the following
intriguing problem: Suppose we have prescribed forms of both
$V_\rmext(\rv)$ and of $\rho(\rv)$, at given $\mu,\beta$, how do we
find the specific pair potential $\phi(r)$ that realizes the target
assuming the interparticle potential is described solely by two-body
contributions.  To accomplish this task, we use Newton's method to
solve the Euler-Lagrange equation~\eqref{EQEulerLagrangeMeta} in the
form $c_1(\rv,[\rho,\beta\phi])-\ln\rho(\rv)-\beta V_\rmext(\rv)+\beta
\mu = 0$.  The required `slope' is the metadirect correlation
functional obtained by automatic differentiation keeping $\rho(\rv)$
fixed: $c_\phi(\rv,r';[\rho,\beta\phi])=\delta
c_1(\rv;[\rho,\beta\phi])/ \delta \beta\phi(r')$; see Appendix~B for
its relevance in analyzing fluctuations via the meta-Ornstein-Zernike
route. The results for $\beta\phi(r)$ in Fig.~\ref{FIG3} demonstrate
that the true pair potential can be reconstructed reliably. Some weak
noise is visible in the four $\beta\phi(r)$. Such artifacts tend to
increase for more extreme cases of~$\beta\phi(r)$. Overall the quality
of the results is remarkable.

While Henderson's uniqueness theorem is formulated canonically
\cite{henderson1974uniqueness, wang2020jcp}, here we worked in the
grand ensemble and provide a corresponding proof in Appendix~C. Using
the neural metafunctional, we validated uniqueness in test particle
situations (Fig.~\ref{FIG2}). We also found empirically unique
solutions in a much wider class of inhomogeneous systems: the hard
wall pore (Fig.~\ref{FIG3}) is a representative example.  Future work
should address this generalized inversion problem formally.  Regarding
higher-dimensional systems, when working with planar symmetry and
randomizing the external potential $\beta V_\rmext(x)$
\cite{sammueller2023neural, sammueller2023whyNeural,
  sammueller2023neuralTutorial, sammueller2024pairmatching,
  sammueller2024hyperDFT, sammueller2024whyhyperDFT,
  sammueller2024attraction, bui2024neuralrpm} as well as $\beta
\phi(r)$ to generate training data, our setup remains applicable.
Henderson inversion is then formally guaranteed via conversion from
planar to radial symmetry \cite{dijkman2024ml,
  sammueller2024pairmatching, sammueller2023neural,
  sammueller2024attraction}.  Instead of the test particle route to
$g(r)$ one could use the Ornstein-Zernike equation together with the
neural bulk pair direct correlation function
\cite{sammueller2023neural, sammueller2024pairmatching,
  sammueller2024attraction, dijkman2024ml, bui2024neuralrpm}.  It
would also be interesting to relate to exact solutions for bulk
one-dimensional systems \cite{archer2013lmft, archer2017meanField,
  montero2019, montero2024}, test particle sum rules
\cite{gul2024testParticle}, quantum test particle concepts
\cite{mccary2020prlBlueElectron}, liquid integral equation theory
\cite{pihlajamaa2024closures, goodall2021closure}, entropy functionals
\cite{percus1994, nicholson2021}, data-driven approaches
\cite{chang2022inversion}, differentiable simulations
\cite{wang2023differentiable}, experimental work \cite{belldavies2024,
  stones2018jcp, stones2023jcp}, splitting of interparticle
interactions \cite{bui2024neuralrpm, cox2020pnas, bui2024}, and
self-assembly \cite{wassermair2024}.

\smallskip
We thank Josh Robinson and Sophie Hermann for helpful
discussions. R.~E. acknowledges support of the Leverhulme Trust, grant
no.~EM-2020-029/4. This work is supported by the DFG (Deutsche
Forschungsgemeinschaft) under project no.~551294732.


\begin{thebibliography}{10}

\bibitem{evans1979}
R. Evans, {The nature of the liquid-vapour interface and other topics in the
  statistical mechanics of non-uniform, classical fluids},
  \href{https://doi.org/10.1080/00018737900101365} {Adv. Phys. {\bf 28}, 143
  (1979).}

\bibitem{evans1992}
R. Evans, {Density functionals in the theory of nonuniform fluids},
  \href{https://books.google.de/books?hl=de&lr=&id=-fNr2a4v3bYC&oi=fnd&pg=PA85}
  {Chap.~3 in {\it Fundamentals of Inhomogeneous Fluids}, edited by D.
  Henderson (Dekker, New York, 1992).}

\bibitem{hansen2013}
J.~P. Hansen and I.~R. McDonald, {\it Theory of Simple Liquids}, 4th ed.\
  (Academic Press, London, 2013).

\bibitem{evans2016}
R. Evans, M. Oettel, R. Roth, and G. Kahl, {New developments in classical
  density functional theory},
  \href{https://doi.org/10.1088/0953-8984/28/24/240401} {J. Phys.: Condens.
  Matter {\bf 28}, 240401 (2016).}

\bibitem{schmidt2022rmp}
M. Schmidt, {Power functional theory for many-body dynamics},
  \href{https://doi.org/10.1103/RevModPhys.94.015007} {Rev. Mod. Phys. {\bf
  94}, 015007 (2022).}

\bibitem{rosenfeld1988}
Y. Rosenfeld, Scaled field particle theory of the structure and the
  thermodynamics of isotropic hard particle fluids,
  \href{https://doi.org/10.1063/1.454810} {J. Chem. Phys. {\bf 89}, 4272
  (1988).}

\bibitem{rosenfeld1989}
Y. Rosenfeld, {Free-energy model for the inhomogeneous hard-sphere fluid
  mixture and density-functional theory of freezing},
  \href{https://doi.org/10.1103/PhysRevLett.63.980} {Phys. Rev. Lett. {\bf 63},
  980 (1989).}

\bibitem{tarazona2000}
P. Tarazona, Density functional for hard sphere crystals: a fundamental measure
  approach, \href{https://doi.org/10.1103/PhysRevLett.84.694} {Phys. Rev. Lett.
  {\bf 84}, 694 (2000).}

\bibitem{roth2010}
R. Roth, {Fundamental measure theory for hard-sphere mixtures: a review},
  \href{https://doi.org/10.1088/0953-8984/22/6/063102} {J. Phys.: Condens.
  Matter \textbf{22}, 063102 (2010)}.

\bibitem{teixera2014}
T. Santos-Silva, P. I. C. Teixeira, C. Anquetil-Deck, and D. J. Cleaver,
  {Neural-network approach to modeling liquid crystals in complex confinement,}
  \href{https://doi.org/10.1103/PhysRevE.89.053316} {Phys. Rev. E {\bf 89},
  053316 (2014).}

\bibitem{lin2019ml}
S.-C. Lin and M. Oettel, {A classical density functional from machine learning
  and a convolutional neural network,}
  \href{http://dx.doi.org/10.21468/SciPostPhys.6.2.025} {SciPost Phys. {\bf 6},
  025 (2019).}

\bibitem{lin2020ml}
S.-C. Lin, G. Martius, and M. Oettel, {Analytical classical density functionals
  from an equation learning network,} \href{https://doi.org/10.1063/1.5135919}
  {J. Chem. Phys. {\bf 152}, 021102 (2020).}

\bibitem{cats2022ml}
P. Cats, S. Kuipers, S. de Wind, R. van Damme, G. M. Coli, M. Dijkstra, and R.
  van Roij, {Machine-learning free-energy functionals using density profiles
  from simulations,} \href{https://doi.org/10.1063/5.0042558} {APL Mater. {\bf
  9}, 031109 (2021).}

\bibitem{qiao2020}
C. Qiao, X. Yu, X. Song, T. Zhao, X. Xu, S. Zhao, and K. E. Gubbins, Enhancing
  gas solubility in nanopores: a combined study using classical density
  functional theory and machine learning,
  \href{https://doi.org/10.1021/acs.langmuir.0c01160} {Langmuir {\bf 36}, 8527
  (2020).}

\bibitem{yatsyshin2022}
P. Yatsyshin, S. Kalliadasis, and A. B. Duncan, {Physics-constrained Bayesian
  inference of state functions in classical density-functional theory,}
  \href{https://doi.org/10.1063/5.0071629} {J. Chem. Phys. {\bf 156}, 074105
  (2022).}

\bibitem{malpica-morales2023}
A. Malpica-Morales, P. Yatsyshin, M. A. Duran-Olivencia, and S. Kalliadasis,
  {Physics-informed Bayesian inference of external potentials in classical
  density functional theory,} \href{https://doi.org/10.1063/5.0146920} {J.
  Chem. Phys. {\bf 159}, 104109 (2023).}

\bibitem{fang2022}
X. Fang, M. Gu and J. Wu, {Reliable emulation of complex functionals by active
  learning with error control,} \href{https://doi.org/10.1063/5.0121805} {J.
  Chem. Phys. {\bf 157}, 214109 (2022).}

\bibitem{dijkman2024ml}
J. Dijkman, M. Dijkstra, R. van Roij, M. Welling, J.-W. van de Meent, and B.
  Ensing, Learning neural free-energy functionals with pair-correlation
  matching, \href{https://arxiv.org/abs/2403.15007} {arXiv:2403.15007.}

\bibitem{kelley2024ml}
M. M. Kelley, J. Quinton, K. Fazel, N. Karimitari, C. Sutton, R. Sundararaman,
  Bridging electronic and classical density-functional theory using universal
  machine-learned functional approximations,
  \href{https://doi.org/10.1063/5.0223792} {J. Chem. Phys. {\bf 161}, 144101
  (2024).}

\bibitem{delasheras2023perspective}
D. de las Heras, T. Zimmermann, F. Samm\"uller, S. Hermann, and M. Schmidt,
  Perspective: How to overcome dynamical density functional theory,
  \href{https://doi.org/10.1088/1361-648X/accb33} {J. Phys.: Condens. Matter
  {\bf 35}, 271501 (2023); (Invited Perspective)}.

\bibitem{sammueller2023neural}
F. Samm\"uller, S. Hermann, D. de las Heras, and M. Schmidt, {Neural functional
  theory for inhomogeneous fluids: Fundamentals and applications},
  \href{https://doi.org/10.1073/pnas.2312484120} {Proc. Natl. Acad. Sci. {\bf
  120}, e2312484120 (2023).}

\bibitem{sammueller2023whyNeural}
F. Samm\"uller, S. Hermann, and M. Schmidt, {Why neural functionals suit
  statistical mechanics}, \href{https://doi.org/10.1088/1361-648X/ad326f} {J.
  Phys.: Condens. Matter {\bf 36}, 243002 (2024); (Topical Review)}.

\bibitem{sammueller2023neuralTutorial}
F. Samm\"uller, {Neural functional theory for inhomogeneous fluids --
  Tutorial}; for online access see the URL:
  \href{https://github.com/sfalmo/NeuralDFT-Tutorial}
  {https://github.com/sfalmo/NeuralDFT-Tutorial }.

\bibitem{sammueller2024hyperDFT}
F. Samm\"uller, S. Robitschko, S. Hermann, and M. Schmidt, Hyperdensity
  functional theory of soft matter,
  \href{https://doi.org/10.1103/PhysRevLett.133.098201} {Phys. Rev. Lett. {\bf
  133}, 098201 (2024); Editors' Suggestion.}

\bibitem{sammueller2024whyhyperDFT}
F. Samm\"uller and M. Schmidt, Why hyperdensity functionals describe any
  equilibrium observable, \href{https://doi.org/10.1088/1361-648X/ad98da} {J.
  Phys.: Condens. Matter {\bf 37}, 083001 (2025); (Topical Review).}

\bibitem{sammueller2024pairmatching}
F. Samm\"uller and M. Schmidt, Neural density functionals: Local learning and
  pair-correlation matching,
  \href{https://doi.org/10.1103/PhysRevE.110.L032601} {Phys. Rev. E {\bf 110},
  L032601 (2024); (Letter, Editors' Suggestion).}

\bibitem{sammueller2024attraction}
F. Samm\"uller, M. Schmidt, and R. Evans, Neural density functional theory of
  liquid-gas phase coexistence,
  \href{https://journals.aps.org/prx/accepted/0e073K26I201d705393b0f29fa179ae5f0bd76d28}{Phys.
  Rev. X {(accepted)}; \href{https://doi.org/10.48550/arXiv.2408.15835}
  {arXiv:2408.15835}.}

\bibitem{bui2024neuralrpm}
A. T. Bui and S. J. Cox, Learning classical density functionals for ionic
  fluids, \href{https://doi.org/10.48550/arXiv.2410.02556} {arXiv:2410.02556.}

\bibitem{baydin2018autodiff}
A. G. Baydin, B. A. Pearlmutter, A. A. Radul, and J. M. Siskind, Automatic
  differentiation in machine learning: A survey,
  \href{http://jmlr.org/papers/v18/17-468.html} {J. Machine Learning Res. {\bf
  18}, 1 (2018).}

\bibitem{stierle2024autodiff}
R. Stierle, G. Bauer, N. Thiele, B. Bursik, P. Rehner, and J. Gross, Classical
  density functional theory in three dimensions with GPU-accelerated automatic
  differentiation: Computational performance analysis using the example of
  adsorption in covalent-organic frameworks,
  \href{https://doi.org/10.1016/j.ces.2024.120380} {Chem. Eng. Sci. {\bf 298},
  120380 (2024).}

\bibitem{henderson1974uniqueness}
R. L. Henderson, A uniqueness theorem for fluid pair correlation functions,
  \href{https://doi.org/10.1016/0375-9601(74)90847-0} {Phys. Lett. A {\bf 49},
  197 (1974).}

\bibitem{johnson1964}
M. D. Johnson, P. Hutchinson, and N. H. March, Ion-ion oscillatory potentials
  in liquid metals \href{https://doi.org/10.1098/rspa.1964.0233} {Proc. Roy.
  Soc. A {\bf 282}, 283 (1964).}

\bibitem{levesque1985}
D. Levesque, J. J. Weis, and L. Reatto, Pair interaction from structural data
  for dense classical liquids,
  \href{https://doi.org/10.1103/PhysRevLett.54.451} {Phys. Rev. Lett. {\bf 54},
  451 (1985).}

\bibitem{enderby1988}
J. E. Enderby, The influence of interatomic forces on the structure of liquids:
  An experimental approach, \href{https://doi.org/10.1080/01418618808205170}
  {Phil. Mag. A {\bf 58}, 5 (1988).}

\bibitem{evans1990reverseMC}
R. Evans, {Comment on reverse Monte Carlo simulation},
  \href{https://doi.org/10.1080/08927029008022403} {Molec. Sim. {\bf 4}, 409
  (1990).}

\bibitem{percus1986}
J. K. Percus, Some solvable models of nonuniform classical fluids,
  \href{https://doi.org/10.1007/BF01010452} {J. Stat. Phys. {\bf 42}, 921
  (1986).}

\bibitem{mcgreevy2001}
R. L. McGreevy, Reverse Monte Carlo modelling,
  \href{https://doi.org/10.1088/0953-8984/13/46/201} {J. Phys.: Condens. Matter
  {\bf 13}, R877 (2001).}

\bibitem{frommer2019}
F. Frommer, M. Hanke, and S. Jansen, A note on the uniqueness result for the
  inverse Henderson problem, \href{https://doi.org/10.1063/1.5112137} {J. Math.
  Phys. {\bf 60}, 093303 (2019).}

\bibitem{frommer2024}
F. Frommer and M. Hanke, A variational framework for the inverse Henderson
  problem of statistical mechanics,
  \href{https://doi.org/10.1007/s11005-022-01563-w} {Lett. Math. Phys. {\bf
  112}, 71 (2022).}

\bibitem{moradzadeh2021deeplearning}
A. Moradzadeh and N. R. Aluru, Understanding simple liquids through statistical
  and deep learning approaches, \href{https://doi.org/10.1063/5.0046226} {J.
  Chem. Phys. {\bf 154}, 204503 (2021).}

\bibitem{toth2005}
G. T\'oth, N. Kir\'aly, and A. Vrabecz, Pair potentials from diffraction data
  on liquids: A neural network solution,
  \href{http://dx.doi.org/10.1063/1.2102887} {J. Chem. Phys. {\bf 123}, 174109
  (2005).}

\bibitem{wang2020jcp}
H. Wang, F. H. Stillinger, and S. Torquato, Sensitivity of pair statistics on
  pair potentials in many-body systems,
  \href{https://doi.org/10.1063/5.0021475} {J. Chem. Phys. {\bf 153}, 124106
  (2020).}

\bibitem{wang2022pre}
S. Torquato and H. Wang, Precise determination of pair interactions from pair
  statistics of many-body systems in and out of equilibrium,
  \href{https://doi.org/10.1103/PhysRevE.106.044122} {Phys. Rev. E {\bf 106},
  044122 (2022).}

\bibitem{jadrich2015}
R. B. Jadrich, J. A. Bollinger, B. A. Lindquist and T. M. Truskett, Equilibrium
  cluster fluids: pair interactions via inverse design,
  \href{https://doi.org/10.1039/C5SM01832C} {Soft Matter {\bf 11}, 9342
  (2015).}

\bibitem{wang2024jcp}
H. Wang and S. Torquato, Designer pair statistics of disordered many-particle
  systems with novel properties, \href{https://doi.org/10.1063/5.0189769} {J.
  Chem. Phys. {\bf 160}, 044911 (2024).}

\bibitem{dyre2016}
J. C. Dyre, Simple liquids' quasiuniversality and the hard-sphere paradigm,
  \href{https://doi.org/10.1088/0953-8984/28/32/323001} {J. Phys.: Condens.
  Matter {\bf 28}, 323001 (2016).}

\bibitem{stones2019prl}
A. E. Stones, R. P. A. Dullens, and D. G. A. L. Aarts, Model-free measurement
  of the pair potential in colloidal fluids using optical microscopy,
  \href{https://doi.org/10.1103/PhysRevLett.123.098002} {Phys. Rev. Lett. {\bf
  123}, 098002 (2019).}

\bibitem{levy1979}
M. Levy, Universal variational functionals of electron densities, first-order
  density matrices, and natural spin-orbitals and solution of the
  v-representability problem, \href{http://dx.doi.org/10.1073/pnas.76.12.6062}
  {Proc. Natl. Acad. Sci. {\bf 76}, 6062 (1979).}

\bibitem{dwandaru2011}
W. S. B. Dwandaru and M. Schmidt, Variational principle of classical density
  functional theory via Levy's constrained search method,
  \href{http://dx.doi.org/10.1103/PhysRevE.83.061133} {Phys. Rev. E {\bf 83},
  061133 (2011).}

\bibitem{mermin1965}
N. D. Mermin, Thermal properties of the inhomogeneous electron gas,
  \href{https://doi.org/10.1103/PhysRev.137.A1441} {Phys. Rev. {\bf 137}, A1441
  (1965).}

\bibitem{kampa2024metaData}
Code, simulation data, and neural functionals available at:
  \href{https://github.com/S-K-acc/metafunctional}
  {https://github.com/S-K-acc/metafunctional}.

\bibitem{percus1962}
J. K. Percus, {Approximation methods in classical statistical mechanics,}
  \href{https://dx.doi.org/10.1103/PhysRevLett.8.462} {Phys. Rev. Lett. {\bf
  8}, 462 (1962)}.

\bibitem{percus1976}
J. K. Percus, Equilibrium state of a classical fluid of hard rods in an
  external field, \href{https://doi.org/10.1007/BF01020803} {J. Stat. Phys.
  {\bf 15}, 505 (1976).}

\bibitem{robledo1981}
A. Robledo and C. Varea, On the relationship between the density functional
  formalism and the potential distribution theory for nonuniform fluids,
  \href{https://doi.org/10.1007/BF01011432} {J. Stat. Phys. {\bf 26}, 513
  (1981).}

\bibitem{montero2019}
A. M. Montero and A. Santos, Triangle-well and ramp interactions in
  one-dimensional fluids: a fully analytic exact solution,
  \href{https://doi.org/10.1007/s10955-019-02255-x} {J. Stat. Phys. {\bf 175},
  269 (2019).}

\bibitem{tonks1936}
L. Tonks, The complete equation of state of one-, two- and three-dimensional
  gases of hard elastic spheres, \href{https://doi.org/10.1103/PhysRev.50.955}
  {Phys. Rev. {\bf 50}, 955 (1936).}

\bibitem{archer2013lmft}
A. J. Archer, and R. Evans, Relationship between local molecular field theory
  and density functional theory for nonuniform liquids,
  \href{https://doi.org/10.1063/1.4771976} {J. Chem. Phys. {\bf 138}, 014502
  (2013).}

\bibitem{archer2017meanField}
A. J. Archer, B. Chacko, and R. Evans, The standard mean-field treatment of
  inter-particle attraction in classical DFT is better than one might expect,
  \href{https://doi.org/10.1063/1.4993175} {J. Chem. Phys. {\bf 147}, 034501
  (2017).}

\bibitem{montero2024}
A. M. Montero, A. Rodriguez-Rivas, S. B. Yustea, A. Santosa, and M. Lopez de
  Haro, On a conjecture concerning the Fisher-Widom line and the line of
  vanishing excess isothermal compressibility in simple fluids,
  \href{https://doi.org/10.1080/00268976.2024.2357270} {Mol. Phys. e2357270
  (2024); \href{https://doi.org/10.48550/arXiv.2404.02520}{arXiv:2404.02520}.}

\bibitem{gul2024testParticle}
M. G\"ul, R. Roth, and R. Evans, Using test particle sum rules to construct
  accurate functionals in classical density functional theory,
  \href{https://doi.org/10.48550/arXiv.2409.01750} {arXiv:2409.01750.}

\bibitem{mccary2020prlBlueElectron}
R. J. McCarty, D. Perchak, R. Pederson, R. Evans, Y. Qiu, S. R. White, and K.
  Burke, Bypassing the energy functional in density functional theory: direct
  calculation of electronic energies from conditional probability densities,
  \href{https://doi.org/10.1103/PhysRevLett.125.266401} {Phys. Rev. Lett. {\bf
  125}, 266401 (2020).}

\bibitem{pihlajamaa2024closures}
I. Pihlajamaa and L. M. C. Janssen, Comparison of integral equation theories of
  the liquid state, \href{https://doi.org/10.1103/PhysRevE.110.044608} {Phys.
  Rev. E {\bf 110}, 044608 (2024).}

\bibitem{goodall2021closure}
R. E. A. Goodall and A. A. Lee, Data-driven approximations to the bridge
  function yield improved closures for the Ornstein-Zernike equation,
  \href{https://doi.org/10.1039/D1SM00402F} {Soft Matter {\bf 17}, 5393
  (2021).}

\bibitem{percus1994}
J. K. Percus, The structure of density functionals,
  \href{https://doi.org/10.1088/0953-8984/6/23A/015} {J. Phys.: Condens. Matter
  {\bf 6}, A125 (1994).}

\bibitem{nicholson2021}
D. M. Nicholson, C. Y. Gao, M. T. McDonnell, C. C. Sluss, and David J. Keffer,
  Entropy pair functional theory: direct entropy evaluation spanning phase
  transitions, \href{https://doi.org/10.3390/e23020234} {Entropy {\bf 23}, 234
  (2021).}

\bibitem{chang2022inversion}
M.-C. Chang, C.-H. Tung, S.-Y. Chang, J. M. Carrillo, Y. Wang, B. G. Sumpter,
  G.-R. Huang, C. D, and W.-R. Chen, A machine learning inversion scheme for
  determining interaction from scattering,
  \href{https://doi.org/10.1038/s42005-021-00778-y} {Commun. Phys. {\bf 5}, 46
  (2022).}

\bibitem{wang2023differentiable}
W. Wang, Z. Wu, J. C. B. Dietschreit, R. Gomez-Bombarelli, Learning pair
  potentials using differentiable simulations,
  \href{https://doi.org/10.1063/5.0126475} { J. Chem. Phys. {\bf 158}, 044113
  (2023).}

\bibitem{belldavies2024}
M. C. R. Bell-Davies, J. Codina, A. Curran, J. Dobnikar, R. P. A. Dullens, I.
  Pagonabarraga, Direct measurement of repulsive and attractive pair potentials
  using pairs of optical traps, \href{https://doi.org/10.1063/5.0184292} {J.
  Chem. Phys. {\bf 160}, 184201 (2024).}

\bibitem{stones2018jcp}
A. E. Stones, R. P. A. Dullens, and D. G. A. L. Aarts, Communication: Contact
  values of pair distribution functions in colloidal hard disks by
  test-particle insertion, \href{https://doi.org/10.1063/1.5038668} {J. Chem.
  Phys. {\bf 148}, 241102 (2018).}

\bibitem{stones2023jcp}
A. E. Stones and D. G. A. L. Aarts, Measuring many-body distribution functions
  in fluids using test-particle insertion,
  \href{https://doi.org/10.1063/5.0172664} {J. Chem. Phys. {\bf 159}, 194502
  (2023).}

\bibitem{cox2020pnas}
S. J. Cox, Dielectric response with short-ranged electrostatics,
  \href{https://doi.org/10.1073/pnas.2005847117} {Proc. Natl. Acad. Sci. {\bf
  117}, 19746 (2020).}

\bibitem{bui2024}
A. T. Bui and S. J. Cox, A classical density functional theory for solvation
  across length scales, \href{https://doi.org/10.1063/5.0223750} {J. Chem.
  Phys. {\bf 161}, 104103 (2024).}

\bibitem{wassermair2024}
M. Wassermair, G. Kahl, R. Roth, A. J. Archer, Fingerprints of ordered
  self-assembled structures in the liquid phase of a hard-core, square-shoulder
  system, \href{https://doi.org/10.1063/5.0226954} {J. Chem. Phys. {\bf 161},
  124503 (2024).}

\bibitem{evans2015jpcm}
R. Evans and M. C. Stewart, {The local compressibility of liquids near
  non-adsorbing substrates: a useful measure of solvophobicity and
  hydrophobicity?}, \href{https://doi.org/10.1088/0953-8984/27/19/194111} {J.
  Phys.: Condens. Matter {\bf 27}, 194111 (2015).}

\bibitem{evans2019pnas}
R. Evans, M. C. Stewart, and N. B. Wilding, {A unified description of
  hydrophilic and superhydrophobic surfaces in terms of the wetting and drying
  transitions of liquids}, \href{https://doi.org/10.1073/pnas.1913587116}
  {Proc. Natl. Acad. Sci. {\bf 116}, 23901 (2019).}

\bibitem{coe2022prl}
M. K. Coe, R. Evans, and N. B. Wilding, {Density depletion and enhanced
  fluctuations in water near hydrophobic solutes: identifying the underlying
  physics,} \href{https://doi.org/10.1103/PhysRevLett.128.045501} {Phys. Rev.
  Lett. {\bf 128}, 045501 (2022).}

\bibitem{eckert2020}
T. Eckert, N. C. X. Stuhlm\"uller, F. Samm\"uller, and M. Schmidt, Fluctuation
  profiles in inhomogeneous fluids,
  \href{https://doi.org/10.1103/PhysRevLett.125.268004} {Phys. Rev. Lett. {\bf
  125}, 268004 (2020).}

\bibitem{eckert2023fluctuation}
T. Eckert, N. C. X. Stuhlm\"uller, F. Samm\"uller, and M. Schmidt, Local
  measures of fluctuations in inhomogeneous liquids: Statistical mechanics and
  illustrative applications, \href{https://doi.org/10.1088/1361-648X/ace50c}
  {J. Phys.: Condens. Matter {\bf 35}, 425102 (2023).}

\end{thebibliography}

\bigskip\bigskip\bigskip
{\bf \large End Matter}\\

\begin{figure}[!hbt]
  \vspace{1mm}
  \includegraphics[width=.99\columnwidth]{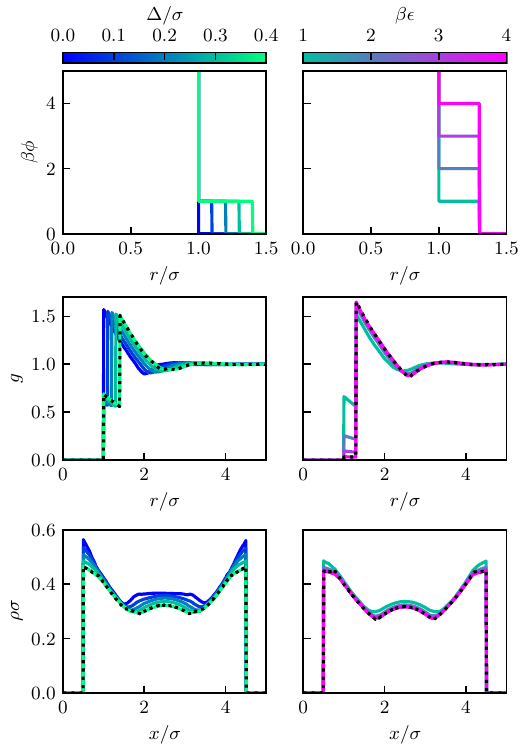}
  \caption{Parametric metadensity functional applications. The scaled
    range ($\Delta/\sigma$) and depth ($\beta\epsilon$) parameters of
    the square shoulder potential are treated as neural network input
    nodes that control the form of $\beta\phi(r)$ (first row).  Percus
    test particle minimization \eqref{EQEulerLagrangeTestParticle}
    yields results for $g(r)$ (second row) for the fluids described by
    the different $\phi(r)$ for $\beta\mu=0$.  Inhomogeneous density
    functional minimization \eqref{EQEulerLagrangeMeta} yields
    corresponding results for $\rho(x)\sigma$ under confinement
    between two hard walls at $x/\sigma=0.5$ and 4.5 (third
    row). Representative simulation results (dotted lines) are shown
    for the largest values of $\Delta/\sigma$ and $\beta\epsilon$
    considered.}
\label{FIG4}
\end{figure}

{\it Appendix A: Parametric metadensity functional.---}Instead of
machine learning the functional dependence on $\beta\phi(r)$ on a
discretized grid, one can work with parametric dependence on the pair
interaction parameters. Although this restricts the general
applicability of the resulting neural metafunctional, it can be more
efficient within the range of applicability. We demonstrate the method
for square shoulder pair potentials with scaled shoulder range
$\Delta/\sigma$ and height $\beta\epsilon$, where both parameters are
used as input nodes, similar to thermal training
\cite{sammueller2024attraction}. Results from inhomogeneous and test
particle DFT of the parameterized metadirect correlation functional
$c_1(\rv;[\rho],\beta \epsilon, \Delta/\sigma)$, shown in
Fig.~\ref{FIG4}, demonstrate high accuracy.  We expect the
parameterization method to apply to many-body interparticle
interaction potentials $u(\rv_1,\ldots,\rv_N;
\alpha_1,\ldots,\alpha_m)$ via machine learning the dependence on $m$
interaction parameters $\alpha_1,\ldots,\alpha_m$ to provide
parameterized access to $c_1(\rv;[\rho,\beta u]) = c_1(\rv;[\rho],
\beta, \alpha_1,\ldots,\alpha_m)$.

\bigskip

\begin{figure}[tb]
  \vspace{1mm}
  \includegraphics[width=.99\columnwidth]{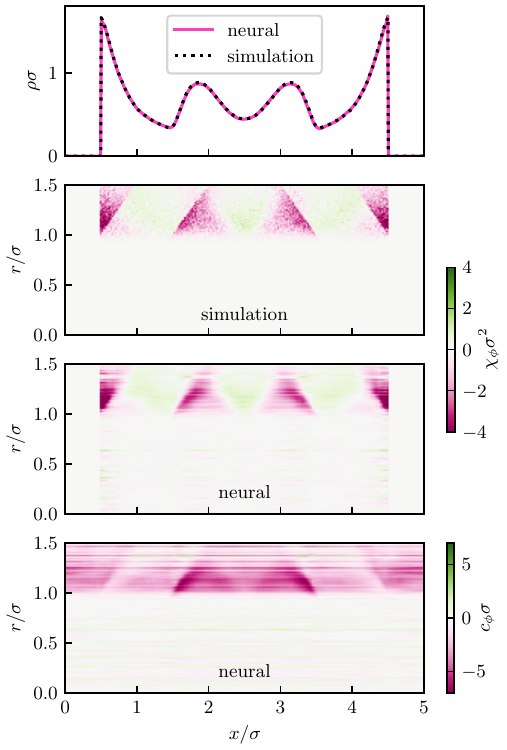}
  \caption{Meta-OZ results. The density profile for $\beta\mu=1$ for
    the truncated LJ system, with $r_c=1.5\sigma$, in hard-wall
    confinement is shown (first panel). Corresponding results are
    shown, as a function of scaled position $x/\sigma$ and
    interparticle distance $r/\sigma$, for the local
    metacompressibility $\chi_\phi(x,r)$ from simulations (second
    panel) and from the neural functional (third panel), and for the
    metadirect correlation function $c_\phi(x,r)$ (fourth panel).}
\label{FIG5}
\end{figure}

{\it Appendix B: Fluctuations.---}We address the fluctuation structure
by functionally differentiating the (logarithm of the) Euler-Lagrange
equation \eqref{EQEulerLagrangeMeta} with respect to $\beta\phi(r')$
keeping $\mu, \beta$ and $V_\rmext(\rv)$ fixed. Re-arranging the
result yields the following exact meta-Ornstein-Zernike (OZ) equation:
\begin{align}
  c_{\phi}(\rv,r';[\rho,&\beta\phi])
  =
  \frac{\chi_\phi(\rv,r')}{\rho(\rv)}\notag\\&
  -\int d\rv'' c_2(\rv,\rv'';[\rho,\beta\phi])\chi_\phi(\rv'',r'),
  \label{EQmetaOrnsteinZernike}
\end{align}
where $c_2(\rv,\rv';[\rho,\beta\phi])=\delta
c_1(\rv;[\rho,\beta\phi])/\delta\rho(\rv')$ \cite{hansen2013,
  evans1992, evans1979} and the local metacompressibility is
$\chi_\phi(\rv,r')=\delta\rho(\rv)/\delta \beta\phi(r')|_{\beta
  (V_\rmext-\mu)}$, generalizing the local compressibility
$\chi_\mu(\rv)$ \cite{evans2015jpcm, evans2019pnas, coe2022prl} and
its extensions \cite{eckert2020, eckert2023fluctuation,
  sammueller2024hyperDFT, sammueller2024whyhyperDFT}.
Equation~\eqref{EQmetaOrnsteinZernike} mirrors closely the
inhomogeneous two-body~\cite{evans1979, evans1992, hansen2013,
  evans2016, schmidt2022rmp}, nonequilibrium \cite{schmidt2022rmp},
fluctuation \cite{eckert2020, eckert2023fluctuation}, and hyperdensity
\cite{sammueller2024hyperDFT, sammueller2024whyhyperDFT} OZ
relations. Figure~\ref{FIG5} shows representative results for
$\chi_\phi(\rv,r')$ and $c_\phi(\rv,r')$ for the LJ system. The highly
structured behaviour in both functions reflects the oscillations in
the density profile.

One can readily show that the local metacompressibility constitutes
the following average: $\chi_\phi(\rv',r)=-\cov(\hat\rho(\rv'), \hat
G(r))$, where the density `operator' is
$\hat\rho(\rv)=\sum_i\delta(\rv-\rv_i)$ and we have defined a global
measure of mutual particle distances $\hat
G(r)=\frac{1}{2}\sum_{ij}'\delta(r-|\rv_i-\rv_j|)$, where the primed
double sum is over distinct pairs $i\neq j$. Then
\eqr{EQmetaOrnsteinZernike} can be viewed as a special case of the
hyper-OZ relation \cite{sammueller2024hyperDFT,
  sammueller2024whyhyperDFT} for the observable $\hat A = -\hat G(r)$.

The test particle situation requires functionally differentiating the
corresponding Euler-Lagrange
equation~\eqref{EQEulerLagrangeTestParticle} with respect to
$\beta\phi(r')$, yielding the test particle meta-OZ relation:
\begin{align}
  c_\phi(r,r';[\rho_g,&\beta\phi])  - \delta(r-r') =
  \frac{\chi_{g}(r,r')}{\rho_g(r)}
  \notag\\&
  -\int d\rv'' c_2(\rv,\rv'';[\rho_g,\beta\phi]) \chi_g(|\rv''|,r'),
  \label{EQmetaOrnsteinZernikeTestParticle}
\end{align}
where $\rho_g(r)=\rho_b g(r)$ and the scaled bulk pair
metacompressibility is $\beta \chi_g(r,r')=\delta
\rho_g(r)/\delta\phi(r')$, which is closely related to $\delta
g(r)/\delta\phi(r')$, as identified and highlighted as important by
Wang {\it et al.}~in their introduction \cite{wang2020jcp}.

\bigskip

{\it Appendix C: Functional relationships.---}As is pertinent to the
treatment in the main text, we give a grand canonical version of
Henderson's theorem. The equilibrium probability density is given by
$f_0 = \e^{-\beta(H-\mu N)}/\Xi$, where the Hamiltonian $H = K + \Phi
+ {\cal V}_\rmext$, with kinetic energy $K=\sum_{i=1}^N\pv_i^2/(2m)$,
interparticle potential energy $\Phi = \sum_{i\neq
  j}\phi(\rv_i-\rv_j)/2$, and external potential energy ${\cal
  V}_\rmext = \sum_{i=1}^N V_\rmext(\rv)$. We keep only the pair
potential $\phi(r)$. Wang {\it et al.}  \cite{wang2020jcp} generalize
$\Phi$ to include higher-body potentials, but work in the canonical
ensemble.  Following Ref.~\cite{hansen2013}, Appendix B, we introduce
the functional
\begin{align}
  \Omega[f] &= \Tr f (H - \mu N + \beta^{-1} \ln f).
\end{align}
In equilibrium: $\Omega[f_0]=-\beta^{-1}\ln \Xi$, where $\Xi=\Tr
\e^{-\beta(H-\mu N)}$ is the grand partition function, and therefore
$\Omega[f_0]=\Omega$ is the grand potential. One can prove
\begin{align}
  \Omega[f] \geq \Omega[f_0]
  \label{EQOmegaInequality}
\end{align}
using a Gibbs-Bogoliubov inequality.

To derive Henderson's theorem, consider two different Hamiltonians --
the original $H$ and $H' = K + \Phi' + \calV'$. Now $\Phi'$ is a sum
of pair potentials $\phi'$, and $\calV_\rmext'$ corresponds to the
external potential $V_\rmext'(\rv)$. The chemical potential $\mu'$
might also be different. Associated with $H'$ are the equilibrium
$f_0'$ and $\Omega_0'$.

Equation \eqref{EQOmegaInequality} asserts: $\Omega' = \Tr
f_0'(H'-\mu'N+\beta^{-1}\ln f_0') \leq \Tr f_0(H'-\mu'N+\beta^{-1}\ln
f_0)$. The right hand side is $\Tr f_0(H - \mu N +
\Phi'-\Phi+\calV_\rmext' - \calV_\rmext + \mu N - \mu' N +
\beta^{-1}\ln f_0)$. Thus
\begin{align}
  \Omega' \leq \Omega[f_0]
  + \Tr f_0 [\Phi' - \Phi 
    +   (\calV_\rmext'-\mu' N)-(\calV_\rmext - \mu N)].
  \label{EQOmegaPrimeInequality1}
\end{align}
Note that $\calV_\rmext - \mu N = \sum_{i=1}^N(V_\rmext(\rv_i)-\mu) =
-\int d\rv\hat\rho(\rv)\psi(\rv)$ with 'intrinsic' chemical potential
$\psi(\rv) = \mu - V_\rmext(\rv)$.  Then \eqr{EQOmegaPrimeInequality1}
reads
\begin{align}
  \Omega' \leq \Omega + \Tr f_0(\Phi'-\Phi)
  - \int d\rv \rho(\rv)[\psi'(\rv)-\psi(\rv)].
  \label{EQOmegaPrimeInequality2}
\end{align}
Swapping primed and unprimed variables, i.e., using $H$ and $f_0$, one
finds
\begin{align}
  \Omega \leq \Omega' + \Tr f_0'(\Phi-\Phi')
  -\int d\rv \rho'(\rv)[\psi(\rv)-\psi'(\rv)],
  \label{EQOmegaPrimeInequality3}
\end{align}
where $\rho(\rv)$ is the equilibrium one-body density corresponding to
$f_0$ and $\rho'(\rv)$ that for $f_0'$.

i) Suppose the pair potentials are identical: $\phi'(r)=\phi(r)$,
i.e., we are considering identical fluids in a reservoir at given
$\mu'=\mu$, but exposed to different external potentials,
$V_\rmext'(\rv)$ and $V_\rmext(\rv)$.  What occurs if one requires
$\rho'(\rv) = \rho(\rv)$? Adding Eqs.~\eqref{EQOmegaPrimeInequality2}
and \eqref{EQOmegaPrimeInequality3}:
\begin{align}
  \Omega' + \Omega \leq \Omega + \Omega',
  \label{EQOmegaContradiction}
\end{align}
a clear contradiction as equality holds only in the trivial case of
the primed and unprimed systems being identical, i.e., there is a
unique $V_\rmext(\rv)$ [or $\psi(\rv)$] that gives rise to a given
$\rho(\rv)$, a key result in the foundation of DFT \cite{evans1979,
  hansen2013, mermin1965}.

ii) Suppose now $\psi'(\rv) = \psi(\rv)$, i.e.\ the intrinsic chemical
potentials are identical, but the pair potentials can be different.
The final terms in Eqs.~\eqref{EQOmegaPrimeInequality2} and
\eqref{EQOmegaPrimeInequality3} vanish, and one can write
Eqs.~\eqref{EQOmegaPrimeInequality2} and
\eqref{EQOmegaPrimeInequality3} respectively as
\begin{align}
  &\Omega'  \leq \Omega + \frac{1}{2} \int d\rv_1 d\rv_2
  \rho^{(2)}(\rv_1, \rv_2)[\phi'(\rv_{12}) - \phi(\rv_{12})],
  \label{EQOmegaPrimeInequality4}
  \\
  &\Omega \leq \Omega' + \frac{1}{2} \int d\rv_1 d\rv_2
  \rho^{(2)'}\!\!(\rv_1,\rv_2)[\phi(\rv_{12})-\phi'(\rv_{12})],
  \label{EQOmegaPrimeInequality5}
\end{align}
where $\rv_{12}=\rv_1-\rv_2$ and $\rho^{(2)}(\rv_1,\rv_2)$ is the
two-body distribution function corresponding to $f_0$ and
$\rho^{(2)'}(\rv_1,\rv_2)$ that for $f_0'$. Suppose now
$\rho^{(2)}(\rv_1,\rv_2) = \rho^{(2)'}(\rv_1,\rv_2)$ then adding
Eqs.~\eqref{EQOmegaPrimeInequality4} and
\eqref{EQOmegaPrimeInequality5} one finds again
\eqr{EQOmegaContradiction}. The contradiction implies that there is a
unique $\phi(r)$ that gives rise to a given $\rho^{(2)}(\rv_1,\rv_2)$.

This argument is equivalent to that of Henderson
\cite{henderson1974uniqueness} who worked in the canonical ensemble --
see his Eqs.~(4) and (5). For the uniform fluid,
$\rho^{(2)}(\rv_1,\rv_2) = \rho_b^2 g(r_{12};\rho_b)$ where $\rho_b$
is the uniform density at specified $\mu$, $T$ and $g(r_{12};\rho_b)$
is the radial distribution function. It follows that the latter
determines uniquely $\phi(r)$.

\end{document}